\documentclass[twocolumn]{aa}
\usepackage{graphicx}
\usepackage{txfonts}
\usepackage{natbib}
\begin{document}

   \title{Astrophysical water masers: Line profiles analysis}
   \titlerunning{H$_2$O maser line profiles analysis}


   \author{W.H.T. Vlemmings\inst{1}\and
        H.J. van Langevelde\inst{2}
          }

   \offprints{WV (wouter@jb.man.ac.uk)\\ Based on research carried out while WV was based in Leiden.}

   \institute{Department of Astronomy, Cornell University, Ithaca, NY 14853-6801, U.S.A. 
        \and
              Joint Institute for VLBI in Europe, Postbus 2, 
                7990~AA Dwingeloo, the Netherlands}

   \date{Received ; accepted }


\abstract{ The changes in the spectral line profile of the 22~GHz
H$_2$O maser are calculated as a function of emerging maser flux. We
address not only the narrowing and re-broadening of the maser lines,
but also the deviations of Gaussian symmetry as a result of the
various hyperfine components of the 22~GHz maser line. Non-LTE models
of the H$_2$O maser transition, including hyperfine structure and
cross-relaxation, are compared with high spectral resolution Very Long
Baseline Interferometry (VLBI) observations of the H$_2$O masers in
the circumstellar envelopes (CSEs) of a sample of late-type
stars. This yields estimates on the thermal width in the maser region
as well as the emerging maser flux and thus the level of
saturation. We also discuss the effect of a velocity gradient along
the maser path on the line widths and shapes of the line profile 
and the effect of the geometry of the maser region. We find that the
typical velocity shift along the maser path is of the order of $1.0$
km~s$^{-1}$. The effect of this shift on the shape of the maser
spectrum is difficult to distinguish from the effect of the hyperfine
components.}

   \maketitle

\section{Introduction}

 The spectral line profiles of astrophysical masers contain important
information on the conditions of the masing region, such as the
thermal line width of the particle distribution and the optical
depth. However, the degree of saturation of the maser has a strong
influence on the maser profile and cannot be determined directly. It
has been shown before, that a detailed analysis of the maser line
widths can provide some information on the maser saturation
(e.g. Goldreich \& Kwan 1974; Litvak 1970). The line profiles first
get narrower with increasing radiative flux and then re-broaden when
the maser becomes radiatively saturated. However, if there is an
effective redistribution of the populations of the excited states
due to cross-relaxation, the re-broadening will only occur until the
radiative flux is large enough that the rate of stimulated emission
exceeds the rate for cross-relaxation. Additionally, in the case of
the 22 GHz rotational H$_2$O maser transition ($6_{16} - 5_{23}$), the
interaction between the different hyperfine components ($F=7-6, 6-5,
5-4, 5-6, 6-6$ and $5-5$) complicates a straightforward line
analysis. Several papers have addressed this issue (e.g. Nedoluha \&
Watson 1991). The manifestation of the hyperfine components in the observed
maser spectra also depends on the maser thermal line width, as well as
the saturation and beaming angle. This is due to the fact that the
hyperfine components cause significant deviations from Gaussian
symmetry. The deviations can be classified by a skewness parameter. In
this paper we examine the feasibility of using the skewness of the
profile together with the line widths to disentangle the thermal line
width of the 22~GHz masing medium and the level of saturation and
beaming angle. We examine a sample of late-type stars and compare
their high resolution total intensity spectra to our non-LTE models.

The H$_2$O masers around late-type stars are thought to occur in the
expanding wind at a few hundred AU from the central star (e.g. Lane et
al. 1987). The temperature of the masing gas is expected to be less
than $1000$~K, according to the analysis of H$_2$O maser pumping
schemes by Neufeld \& Melnick (1990). As a result the intrinsic,
thermal line width of the maser should be $\la
1.5$~km~s$^{-1}$. Larger line widths can be caused by turbulence of
the masing gas. Model estimates of the intrinsic thermal line widths
can thus give an upper limit to the temperature in the maser region
and to the effect of turbulence. At a few hundred AU from the central
star the outflowing material is still being accelerated. Therefore, as
shown in Rosen et al.\ (1978), the velocity coherent paths in the
tangential and in the radial direction are of similar length. As a
result, for the H$_2$O masers neither radially nor tangentially
beaming necessarily dominates.  The H$_2$O maser spots have typical
sizes of $10^{13}$~cm (Reid \& Moran 1981), but as determined from
VLBI observations they can be as small as a few times $10^{12}$~cm
(Imai et al. 1997). Due to beaming, the actual size of the H$_2$O
maser cloud is generally much larger than the observed size. Goldreich \&
Keeley (1972) have shown that for H$_2$O masers a factor of $50$
difference is likely. The maser path length could thus be as long as
$10^{14} - 10^{15}$~cm. Along this path the velocity can change by a
significant fraction of the intrinsic thermal line width, especially
near the edge of the envelope where the velocity gradient is
steep. Here we also examine the effect of such velocity changes on the
shape of the 22~GHz H$_2$O maser profiles.

\section{Background}

\begin{figure}
\begin{center}
   \resizebox{\hsize}{!}{\includegraphics{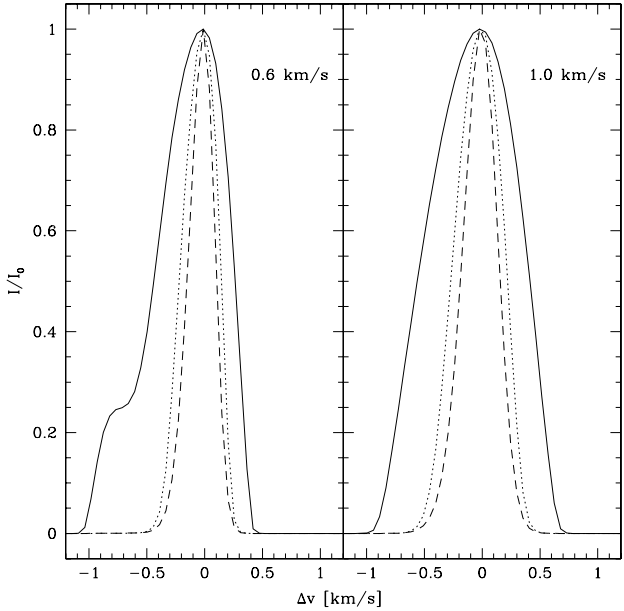}} \hfill
\end{center}
   \caption{Scaled intensity lines profiles for the non-LTE models
   with $v_{\rm th}=0.6$ km~s$^{-1}$ (left) and $v_{\rm th}=1.0$
   km~s$^{-1}$ (right). From inside out, the dashed line has
   emerging maser flux $T_{\rm b}\Delta\Omega = 10^8$, the
   short-dashed line has $T_{\rm b}\Delta\Omega = 10^{10}$ and the
   solid line has $T_{\rm b}\Delta\Omega = 10^{11}$.}
\label{profile}
\end{figure}
\begin{figure}
   \resizebox{\hsize}{!}{\includegraphics{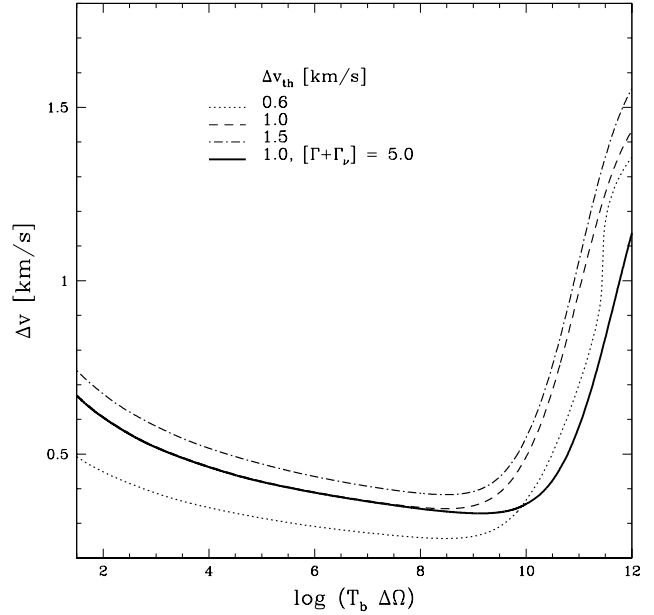}} \hfill
   \caption{Line width as a function of emerging maser flux $T_{\rm
   b}\Delta\Omega$ for models with different thermal width. The thick
   solid line is the model for $v_{\rm th} = 1.0$ km~s$^{-1}$ with [$\Gamma +
   \Gamma_\nu$]$ = 5.0$, the other lines are for [$\Gamma +
   \Gamma_\nu$]$ = 1.0$.}
\label{width}
\end{figure}

 The rate equations for the populations of the upper and lower energy
levels of the various hyperfine lines of the 22 GHz H$_2$O maser have
been solved along a uni-directional maser beam using the method
described in Nedoluha \& Watson (1992, hereafter NWa) and used in
Vlemmings et al. (2002, hereafter V02).  We have included only the
three strongest hyperfine components ($F=7-6, 6-5$ and $5-4$), since
it has previously been shown that the weakest of the hyperfine lines
are negligible (Nedoluha \& Watson 1991, Deguchi \& Watson 1986). In
V02, the rate equations are solved including a magnetic field, thus
including all the different magnetic substates. The magnetic field
does not noticeably influence the line shapes and line widths of the
total intensity maser lines. Here we limit ourselves to the case of no
magnetic field.

For the population levels of the upper states we solve
\begin{eqnarray}
0 & = & \lambda_i(v) - (\Gamma+\Gamma_\nu)n_i(v) + {
R_{ij}}(v)(n_j(v) - n_i(v)) \nonumber \\ 
& & + \phi(v)({\Gamma_\nu \over \sum
g_i})\int dv\sum g_i n_i(v).
\label{eq1_9}
\end{eqnarray}
A similar but reversed equation is solved for the population of the
lower levels.

Here $n_i(v)$ and $n_j(v)$ are the population numbers for respectively
the upper ($6_{\rm 16}$) and the lower ($5_{\rm 23}$) rotational
levels. $i$ denotes the upper level hyperfine states ($F=7,6$ or $5$),
$j$ is the related lower hyperfine state ($F'=6,5$ or $4$). The
population levels are solved as a function of molecular velocity $v$
and at different positions along the maser path. The statistical
weights are designated by $g_i$ and $g_j$, and are from Kukolich
(1969). The pump rate $\lambda_i(v)$ is assumed to be the same for the
different hyperfine components and has a Maxwellian distribution. The
results depend on the the ratio $\Delta\lambda / \lambda_i$ where
$\Delta\lambda$ is the difference between the pump rate into the upper
and lower levels. This ratio is of the order of a few percent
(Anderson \& Watson 1993). The rate for stimulated emission $R_{ij}$
is calculated using the local maser intensity and the hyperfine
interaction coefficients as described in NWa. $\Gamma$ is the decay
rate for the molecular excitations.  The cross-relaxation rate
$\Gamma_\nu$ describes the reabsorbtion of previously emitted infrared
pump photons trapped in optically thick transitions of the maser
system, which generate a newly excited molecule at random, within a
Maxwellian velocity distribution ($\phi(v)$). This has the effect
of postponing the rebroadening of the maser line due to maser
saturation. The rate for redistribution in velocity is not exactly
the same for the redistribution among the different hyperfine
components. However, both are expected to be roughly equal to the
inverse lifetime of the state due to emission of infrared radiation.
In NWa it was shown that the results scale by [${\rm flux}/(\Gamma +
\Gamma_\nu)$]. Most of the calculations are performed for $[\Gamma +
\Gamma_\nu] = 1.0$~s$^{-1}$ although we also investigate higher values
of $\Gamma$ and $\Gamma_\nu$. A similar effect as the
cross-relaxation due to reabsorbtion is produced by elastic collisions
between the maser molecules and intermixed H$_2$ molecules, that
change the velocity but not the energy of the colliding
particles. This effect is described in detail in Elitzur (1990), where
it was shown that due to elastic collisions, the narrowing of the
maser line can be extended significantly. However, the analysis by
Elitzur (1990) indicates that for maser transitions that have large
infrared optical depths, such as the H$_2$O maser discussed here, the
effect of elastic collisions is small compared to that of $\Gamma$ and
$\Gamma_\nu$.

Eq.~\ref{eq1_9} is iteratively solved along the maser
beam, as the maser intensity itself, determined using the radiative
transfer equations in NWa, depends on the level populations.
We assume nearly one-dimensional maser propagation, with the beaming
of the maser radiation represented by the solid angle
$\Delta\Omega$. The emerging maser fluxes will thus be represented in
$T_{\rm b}\Delta\Omega$, with $T_{\rm b}$ the brightness
temperature. The beaming angle $\Delta\Omega$ is not a constant
property throughout the entire maser regime. It depends on the maser
optical depth $|\tau|$ as beaming is more pronounced for the stronger
masers. The beaming angle in the unsaturated cases is roughly
proportional to $|\tau|^{-1}$, while in the saturated case it is
proportional to $|\tau|^{-2}$ (Elitzur 1992). All calculations were
performed with $T_{\rm b}\Delta\Omega = 0.1$ K~sr, for the radiation
incident onto the maser region. It was verified in NWa and V02 that
the results are insensitive to the chosen value.

\begin{figure}
\begin{center}
   \resizebox{\hsize}{!}{\includegraphics{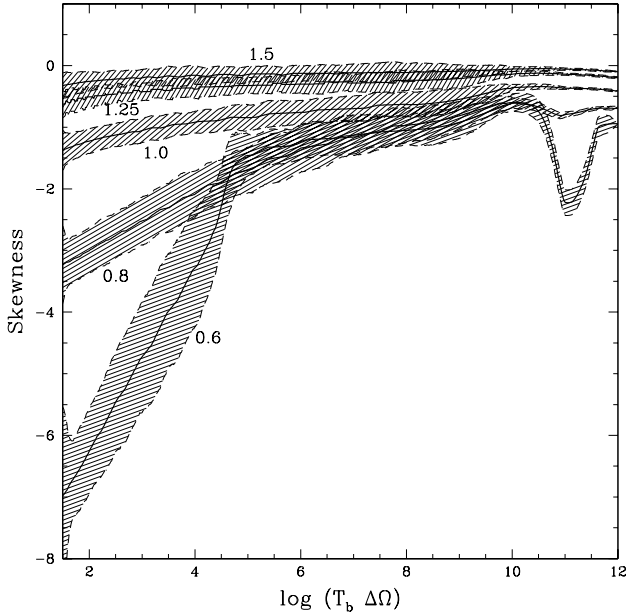}} \hfill
\end{center}
   \caption{Skewness parameter as a function of emerging maser brightness $T_{\rm b}\Delta\Omega$. The lines are labeled for different thermal width. The shaded areas indicate the $3\sigma$ errors when including observational errors as described in the text.}
\label{skew}
\end{figure}

\subsection{Profile shapes}

 In Fig.~\ref{profile}, we show an example of the line profiles
 resulting from our calculations. The example profiles have been
 calculated for two different values of the FWHM thermal width
 ($v_{\rm th}$) of the Maxwellian particle distribution. Assuming
 a kinetic temperature $T$ in Kelvin, $v_{\rm th}\approx
 0.5\times(T/100)^{1/2}$ (NWa). Line profiles are presented for different
 values of emerging flux ($T_{\rm b}\Delta\Omega = 10^8, 10^{10}$ and
 $10^{11}$). Due to the hyperfine interaction the maser profiles show
 clear deviations from Gaussian symmetry with increasing maser
 brightness. The effects are largest for the smallest thermal
 widths. We also notice the re-broadening of the profiles when the
 maser starts to get saturated. This can also be seen in
 Fig.~\ref{width}, which shows the maser line width as a function of
 emerging flux. In the unsaturated case the maser gets increasingly
 narrow, when the masers starts to get saturated the line width
 increases. The figure shows the relations for $[\Gamma +
 \Gamma_\nu]=1$ and $5~{\rm s}^{-1}$. As expected, the maser is able
 to remain unsaturated longer, for higher values of the decay and
 cross-relaxation rate.

 Aside from the narrowing and re-broadening of the maser lines,
Fig.~\ref{profile} shows that~ due to the~ contributions~ of the
hyperfine~ components, the shape of the 22~GHz H$_2$O maser profile is
also a good indicator of the saturation level. The hyperfine
components will cause the profile to be skewed toward the negative
velocities and the skewness will depend on the level of
saturation. The deviation from Gaussian shape can be quantified by the
skewness parameter defined as:

\begin{equation}
{\rm Skewness} = {1 \over \Psi} \int dv [\psi(v) {(v-v_{\rm 0}) \over \sigma}]^3.
\label{skeq}
\end{equation} 

Here $\psi(v)$ is the profile of the maser line, $\sigma$ the standard
deviation and $v_{\rm 0}$ the position of maximum intensity. The parameter
is scaled by $\Psi$, which is the area under the profile. A positive
value of the skewness parameter signifies a distribution with an
asymmetric tail out toward positive velocity, while a negative
skewness parameter indicates a negative tail. Fig.~\ref{skew} shows
the skewness of the maser profile as a function of emerging brightness
for different values of $v_{\rm th}$, as determined from our
models. For the highly unsaturated masers the skewness parameter is
strongly negative due to the multiple hyperfine components. The
profiles become more symmetric when the F=7-6 hyperfine component
starts to dominate. When the maser starts to saturate, at $T_{\rm
b}\Delta\Omega \approx 10^{10}$ for $[\Gamma + \Gamma_\nu] = 1~{\rm
s}^{-1}$, the skewness parameter decreases again, because the F=7-6
hyperfine component is the first to saturate. As a result, the profile
becomes more asymmetric. This is the most pronounced for the lowest
values of $v_{\rm th}$. For $v_{\rm th} < 1.0$ km~s$^{-1}$ the skewness
parameter increases again for $T_{\rm b}\Delta\Omega > 10^{11}$ after
a strong decrease. For the higher thermal velocity widths the skewness
parameter keeps decreasing slightly in the saturated regime.

Because the skewness parameter is the third moment of the maser
profile, it is strongly affected by observational errors. To include
the influence of the errors, we have simulated our observation by
adding Gaussian distributed errors on top of the model profiles. We
have used a signal-to-noise of 100, typical for the weakest of the
observed maser sources. The result is indicated with the shaded areas
in Fig.~\ref{skew}, these correspond to 3$\sigma$ errors. The effect
of the errors is strongest in the unsaturated regime and decreases
when the maser saturates. The spectral resolution of the maser
profiles also affects the errors in the skewness determination. Higher
spectral resolution observations have smaller errors because of a more
accurate determination of the scaling parameter $\Psi$ and velocity
$v_{\rm 0}$ of the peak intensity. Our models were determined with a
spectral resolution of $0.027$ km~s$^{-1}$, similar to the
observations.

\begin{figure}
  \begin{center}
   \resizebox{\hsize}{!}{\includegraphics{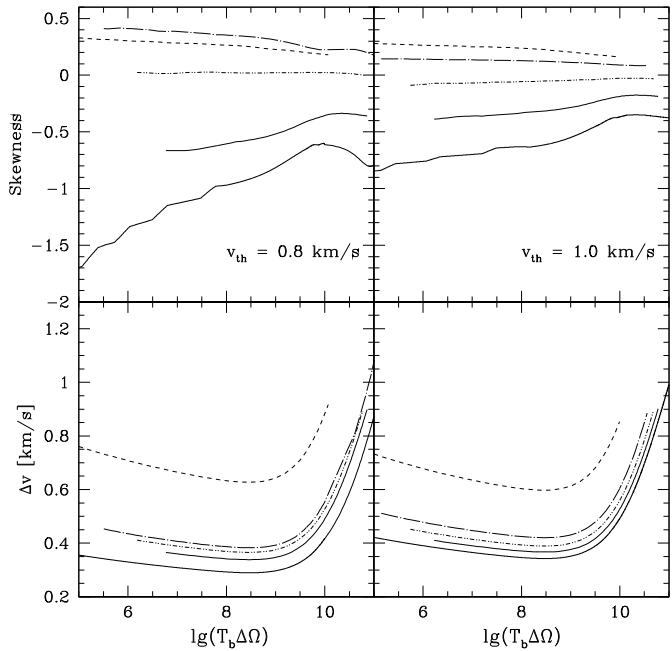}} \hfill
  \end{center}
   \caption{Skewness and line widths for from $v_{\rm th} = 0.8$ (left) and
   $1.0$ km~s$^{-1}$ (right) when including a velocity shift $\Delta V_{\rm m}$
   of $0.4$ (thin solid), $0.6$ (short-dashed), $0.8$ (dashed-dotted) and
   $1.2$ km~s$^{-1}$ (long-dashed) as a function of emerging maser
   brightness $T_{\rm b}\Delta\Omega$. The thick solid line indicates the result
   with constant velocity along the maser path. Due to the
   computational requirements we have not calculated the full
   brightness temperature range for the models which have a velocity
   gradient.}
\label{drift}
\end{figure}

\subsection{Velocity gradients}
 
Obviously the line profile can also be determined by velocity
gradients along the radiation path, giving rise to observable
skewness. Therefore, the calculations above have also been performed
including a shift of velocity along the maser path ($\Delta V_{\rm
m}$). This has been done by replacing the pump rate $\lambda(v)$ in
Eq.~\ref{eq1_9} for all hyperfine components by $\lambda(v+\delta v)$,
where $\delta v$ increases for every step along the maser path. The
effects have been calculated for the models with $v_{\rm th} = 0.8$
and $1.0$~ km~s$^{-1}$. Results for the line widths and skewness
parameter are shown in Fig.~\ref{drift} for $\Delta V_{\rm m}$ of up
to $1.2$ km~s$^{-1}$. In our calculations, the linear velocity
gradient along the maser path is characterized by $\Delta V_{\rm m}$,
the velocity shift between the start and the end of the path. The
small variations on the lines are due to the spectral resolution of
our models, which is equal to the resolution of the observations.

The skewness is very sensitive to $\Delta V_{\rm m}$, especially in
the unsaturated case. The effect of $\Delta V_{\rm m}$ becomes less
for models with a relatively large thermal width. When $\Delta V_{\rm
m}$ becomes significantly larger than the thermal width, the skewness
no longer depends on the intrinsic width. This explains that, in
Fig.~\ref{drift}, the model for $v_{\rm th} = 0.8$ and a
$\Delta V_{\rm m} = 1.2$~km~s$^{-1}$ is less positively skewed than
the model with $\Delta V_{\rm m} = 0.8$~km~s$^{-1}$, while this is not
seen in the models with $v_{\rm th} = 1.0$. Most models show less
deviations of Gaussian symmetry in the saturated case. All the models
in Fig.~\ref{drift} were calculated with a negative velocity gradient
along the maser path.  A gradient in the positive velocity direction
increases the negative skewness of the profiles for all cases. Since
the front side of the envelope has the most negative velocity, the
velocity gradient through the envelope toward the observer is
predominantly negative, for features originating from both sides of
the circumstellar envelope.

We find that a velocity gradient along the maser path increases the
maser line width with an approximately equal amount in both the
unsaturated as well as the saturated case. As soon as $\Delta V_{\rm
m}$ approaches the intrinsic thermal line width, the measured line
widths increase rapidly.  Nedoluha \& Watson (1988) have shown that
for $\Delta V_{\rm m}$ a few times $v_{\rm th}$, the maser line splits
into several distinct, narrow features, each becoming an independent
maser. This has been observed for H$_2$O masers in star-forming
regions (e.g. Genzel et al. 1979, Walker 1984). Our observations of
the circumstellar H$_2$O masers do not show this line splitting,
indicating that the largest velocity shift along the maser path is not
more than $\approx 2.0$ km~s$^{-1}$.

\section{Observations}

\begin{table*}
\caption{Star Sample}
\begin{tabular}{|l|c|cc|c|c|c|c|}
\hline
Name & Type & RA & Dec & Distance & Period & V$_{\rm rad}$ & Peak
flux\\
&&($^{h}~^{m}~^{s}$)&($^{\circ}~{'}~{"}$)&(pc)&(days)&(km~s$^{-1}$)&(Jy)\\
\hline
\hline
S Per & Supergiant & 02 22 51.72 & +58 35 11.4 & 1610 & 822 & -38.1 & 76.1\\
U Her & Mira & 16 25 47.4713 & +18 53 32.867 & 277 & 406 & -14.5 & 12.5
\\
NML Cyg & Supergiant & 20 46 25.7 & +40 06 56 & 1220 & 940 & 0.0 & 48.2\\
VY CMa & Supergiant & 07 22 58.3315 & -25 46 03.174 & 1500 & 2000 & 22.0 & 244.1\\
\hline
\end{tabular}
\label{sample}
\end{table*}

The observations were performed at the NRAO\footnote{The National
Radio Astronomy Observatory is a facility of the National Science
Foundation operated under cooperative agreement by Associated
Universities, Inc.}  Very Long Baseline Array (VLBA) at December 13th
1998 as part of the project designed to measure the circular
polarization of circumstellar H$_2$O masers. The results of the
polarization measurements are presented in V02. At the frequency of
the $6_{16}-5_{23}$ rotational transition of H$_{2}$O, 22.235 GHz, the
average beam width is $\approx 0.5 \times 0.5$ mas. This allows us to
resolve the different H$_{2}$O maser features in the CSE. As explained
in V02, the data were correlated twice, once with modest ($7.8$ kHz $=
0.1$~km~s$^{-1}$) spectral resolution, and once with high spectral
resolution ($1.95$ kHz $= 0.027$ km~s$^{-1}$). In this paper we use
the total intensity profiles obtained with the high spectral
resolution data. We have performed 6 hours of observations per
source-calibrator pair. The calibrator was observed for $1~1/2$ hours
in a number of scans equally distributed over the 6 hours. We used 2
filters of 1 MHz width, which were overlapped to get a velocity
coverage of $\approx 22$ km~s$^{-1}$.  This covers most of the
velocity range of the H$_2$O maser.

\subsection{Sample}

We observed 4 late type stars, the supergiants S~Per, VY~CMa and
NML~Cyg and the Mira variable star U~Her. They are listed in
Table.~\ref{sample} with type, position, distance and period. The
sources were selected on the basis of 2 criteria; strong H$_2$O masers
and the availability of SiO and/or OH maser polarization observations.
The peak fluxes measured in the observations are shown in
Table.~\ref{sample}. On the total intensity maps the noise is between
$\approx 0.08 - 0.3$~Jy.

\section{Results}

\begin{table*}
\caption{Results}
\begin{center}
\begin{tabular}{|l|c|c|c|c|c|c|}
\hline
Name & $V_{\rm rad}$ & Flux (I) & $\Delta
v_{\rm L}$ & Skewness & $3\sigma^+_{\rm skew}$ & $3\sigma^-_{\rm skew}$ \\
&(km~s$^{-1}$)&(Jy)&(km~s$^{-1}$)& & & \\
\hline
\hline
S Per 
  & -23.2 & 10.0 & 0.44 & -0.29 & ${+0.11}$ & ${-0.24}$ \\
  & -27.2 & 76.1 & 0.88 & -2.21 & ${+0.31}$ & ${-0.43}$ \\ 
  & -27.2 & 14.2 & 0.51 & 0.04 & ${+0.06}$ & ${-0.10}$ \\
  & -28.9 & 29.7 & 0.61 & 4.35 &${+0.84}$ & ${-0.66}$ \\
  & -30.4 & 15.4 & 0.92 & -0.10 & ${+0.04}$ & ${-0.09}$ \\
  & -30.9 & 57.7 & 0.77 & -0.73$^*$ & ${+0.11}$ & ${-0.16}$ \\
  & -33.9 & 11.4 & 1.21 & -0.59 & ${+0.10}$ & ${-0.15}$ \\
  & -34.2 & 14.9 & 0.48 & -0.20 & ${+0.06}$ & ${-0.14}$ \\  
  & -37.1 & 34.7 & 0.54 & -0.25 & ${+0.05}$ & ${-0.09}$ \\
  & -39.2 & 26.6 & 0.64 & 1.31 & ${+0.27}$ & ${-0.22}$ \\
  & -39.3 & 12.9 & 0.69 & -0.17 & ${+0.06}$ & ${-0.11}$ \\
  & -40.1 & 35.6 & 0.53 & -0.19 & ${+0.04}$ & ${-0.06}$ \\
\hline
VY CMa 
 &  10.6 & 155.5 & 0.67 & -0.35$^*$ & ${+0.05}$ & ${-0.09}$\\
 &  10.8 & 244.1 & 0.66 & -0.53$^*$ & ${+0.08}$ & ${-0.11}$\\
 &  12.9 & 115.2 & 0.74 & -1.19 & ${+0.17}$ & ${-0.25}$ \\
 &  13.7 & 47.9  & 0.83 & -0.11 & ${+0.02}$ & ${-0.04}$ \\
 &  15.0 & 17.8  & 0.94 & 0.01 & ${+0.02}$ & ${-0.05}$ \\
 &  18.3 & 58.5  & 0.80 & -0.36 & ${+0.06}$ & ${-0.08}$ \\
 &  24.6 & 126.2 & 0.60 & 0.07 & ${+0.02}$ & ${-0.03}$ \\
 &  25.2 & 65.6  & 0.79 & 0.34 & ${+0.07}$ & ${-0.06}$ \\
 &  27.1 & 25.3  & 0.88 & 0.22 & ${+0.06}$ & ${-0.06}$ \\
 &  28.8 & 43.1  & 0.61 & -0.32 & ${+0.05}$ & ${-0.06}$ \\
\hline
NML Cyg 
 &  -20.0 & 3.66 & 0.54 & 0.64 & ${+0.32}$ & ${-0.36}$ \\
 &  -20.2 & 5.42 & 0.54 & -0.14 & ${+0.11}$ & ${-0.30}$ \\
 &  -20.4 & 10.5 & 0.40 & -0.09 & ${+0.08}$ & ${-0.19}$ \\
 &  -21.2 & 48.2 & 0.66 & -0.03 & ${+0.02}$ & ${-0.03}$ \\
\hline
U Her 
 & -17.6 & 2.91 & 0.50 & 5.47 & ${+1.65}$ & ${-1.60}$ \\
 & -17.6 & 1.16 & 0.73 & 0.69 & ${+0.55}$ & ${-0.63}$ \\
 & -17.8 & 12.20 & 0.75 & 0.69 & ${+0.18}$ & ${-0.16}$ \\ 
 & -19.2 & 2.46 & 0.47 & -0.63 & ${+0.38}$ & ${-0.93}$ \\
 & -19.2 & 2.07 & 0.51 & -0.14 & ${+0.33}$ & ${-0.76}$ \\
 & -19.3 & 1.36 & 0.51 & -0.27 & ${+0.46}$ & ${-1.14}$ \\
\hline
\multicolumn{7}{l}{$^*$ Bad spectral channels affect skewness determination}
\end{tabular}
\end{center}
\label{results_9}
\end{table*}

\begin{figure}
\begin{center}
   \resizebox{\hsize}{!}{\includegraphics{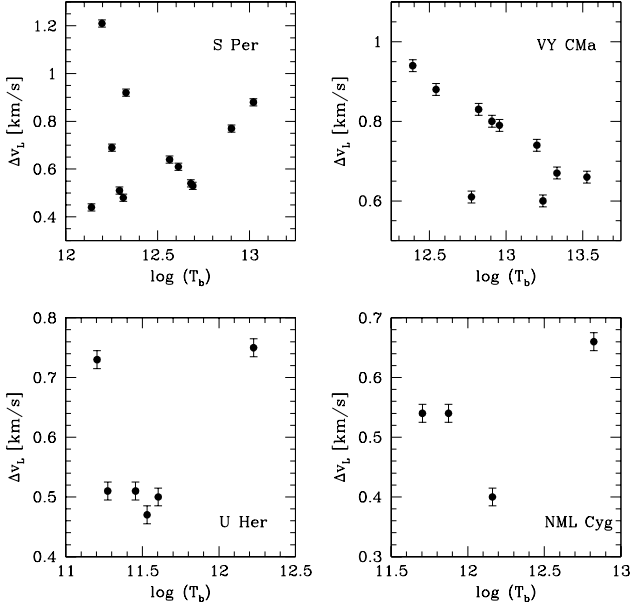}} \hfill
\end{center}
   \caption{Line widths of the observed maser features as a function
   of emerging flux. The error bars are the $3\sigma$ errors.}
\label{wdat}
\end{figure}

\begin{figure}
  \begin{center}
   \resizebox{\hsize}{!}{\includegraphics{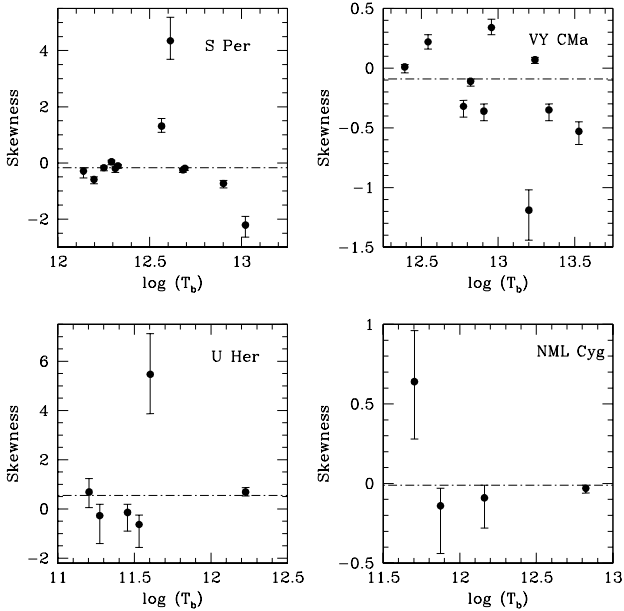}} \hfill
  \end{center}
   \caption{Skewness parameter of the observed maser features as a
   function of emerging flux. The error bars indicate $3\sigma$ errors
   and the dashed-dotted line is the weighted average.}
\label{skewdat}
\end{figure}

 The results of the observations are shown in
Table~\ref{results_9}. The table gives the LSR velocity of the maser
feature, the maser strength in Jy and the line width and skewness
parameter. The features labeled $^*$ have a few bad spectral channels
in the wings of the profile due to interference. This can have a
significant effect on the skewness of the profile; we have verified
that the values listed are the upper limits for these features and the
actual skewness could be up to $\approx 1.0$ lower. The quoted
$3\sigma$ error values of the skewness, are based on the errors in the
determination of $v_{\rm 0}$ and $\sigma$ for use in
Eq.~\ref{skeq}. These depend on the signal-to-noise ratio of the maser
feature, and on the spectral resolution of the observations.

We will plot the observed line width and skewness as a function of
brightness temperature in order to make a comparison with our
models. One should realize that there is not necessarily a single
value for width and velocity shift that is valid for all maser features in
any star. It is natural to expect the results to be quite scattered.

Figs.~\ref{wdat} and \ref{skewdat} show the observed line widths
$\Delta v_L$ and Skewness parameters plotted against the logarithm of
the maser brightness temperature $T_{\rm b}$. This is
calculated using the fluxes listed in Table~\ref{results_9} using the
beam size of the observations as the maser spots are unresolved, thus
$T_{\rm b}$ is formally only a lower limit. However, at the distance
of the star the beam has a diameter of approximately $1$--$2 \times
10^{13}$~cm and is comparable to the typical maser spot size (Reid \&
Moran 1981). We thus expect the determined brightness temperatures to
be 0 with the real values.

\subsection{U~Her}

For the Mira variable star U~Her most of the maser features show line
widths of $\approx~0.5$ km~s$^{-1}$ with 2 of the features showing widths of
$\approx 0.75$ km~s$^{-1}$. There is no clear trend with increasing maser
brightness. The H$_2$O masers around U~Her have also been observed
with the Very Large Array (VLA) by Colomer et al. (2000). They observe
line widths several times wider than found in our observations, but the
VLA is unable to resolve the individual spots and they only used a
spectral resolution of $0.33$ km~s$^{-1}$. It is likely that the larger
line widths are due to the blending of several smaller components.

Because U~Her has the weakest masers, the errors on the skewness
parameter are relatively large. The weighted average of the skewness
is $0.55$, a possible indication of large $\Delta V_{\rm m}$. The
skewness is positive for the maser features with $V_{\rm LSR} \approx
-17.5$ km~s$^{-1}$, while it is negative for the features with $V_{\rm LSR}
\approx -19.0$~km~s$^{-1}$. All the maser features are blue-shifted with
respect to the stellar velocity of $V_{\rm LSR} = -14.4$ km~s$^{-1}$. As
discussed above, $\Delta V_{\rm m}$ will be negative because the
maser exists in a radially accelerating outflow. This results in a
frequency shift along the maser feature in the direction of the
hyperfine components. If the features with a velocity closer to the
stellar velocity are located closer to the star, they will experience
more acceleration along the maser path and thus a larger velocity
gradient. This can explain that the skewness of these features is more
positive than the that of the other features. The maser feature with
the skewness of $5.47$ however, is likely a blend of two overlapping
maser features.

\subsection{NML~Cyg}

The line widths of the maser features around NML~Cyg are found to lie
between $0.4$ and $0.7$ km~s$^{-1}$. Again no clear trend can be determined,
especially since we have only been able to detect 4 maser spots.  MERLIN
observations by Richards et al. (1996) show a much more complex
structure with many features. The line widths they find cover a large
range between $0.4$ and $2.5$ km~s$^{-1}$.

The weighted average of the skewness is $-0.01$. The line widths and
skewness, indicate an intrinsic thermal width of between $0.8$ and
$1.0$~km~s$^{-1}$ and $\Delta V_{\rm m}$ of $\approx
0.75$~km~s$^{-1}$. A larger $\Delta V_{\rm m}$ would result in a more
positive skewness, while a larger thermal width does not allow for the
observed narrow line widths.

\subsection{S~Per}

The line widths of the maser features around S~Per show some
indication of an increase with maser brightness. The line width
increases from $0.4$ to $0.9$ and varies approximately as $\Delta v
\propto (T_{\rm b})^{0.3}$. However, 2 features show a width which is higher
than expected. But, as seen in Fig.~\ref{drift}, this can be due to a
velocity shift along the path which is more than
$1.2$~km~s$^{-1}$. Alternatively, the actual spot size of these
features could be less than $10^{12}$~cm leading to an underestimation
of $T_{\rm b}$. The skewness of the other features indicate a
$\Delta V_{\rm m}$ of not more than $0.8$~km~s$^{-1}$. Similar velocity
shifts have been observed by Richards et al. (1999). The observed line
widths correspond to a thermal line width between $0.8$ and $1.0$
km~s$^{-1}$.

 A combination of both the skewness and line width analysis allows us
to estimate, from our models, the actual emerging maser brightness
temperature and beaming angle. For the estimated velocity shifts
along the maser path and thermal line widths, our models give observed
line widths of $\approx 0.4 - 0.9$ km~s$^{-1}$ for $T_{\rm b}\Delta\Omega$
between $10^{10}$ and $10^{11}$ K. With observed brightness
temperature $T_{\rm b}=10^{12}$--$10^{13}$, this indicates that the
beaming solid angle $\Delta\Omega \approx 10^{-2}$~sr.

\subsection{VY~CMa}

 The line widths of the H$_2$O masers around VY~CMa show a clear
dependence on the brightness temperature. The line width varies with
$\Delta v \propto (T_{\rm b})^{-\alpha}$, where $\alpha \approx 0.3$. A similar
decrease of line width with intensity was observed before on the maser
flare in Cepheus A (Rowland \& Cohen 1986), though there $\alpha
\approx 0.5$.  We observe line widths from $0.6$ to $0.94$
km~s$^{-1}$. MERLIN observations (Richards et al. 1999) have found many
features to have a velocity extent of up to $3$ km~s$^{-1}$, indicating the
presence of large velocity shifts along the maser.

For VY~CMa, the skewness and line widths indicate a thermal width in
the H$_2$O maser region of $\approx 1.0 - 1.2$~km~s$^{-1}$, with
$\Delta V_{\rm m}$ up to $1.25$ km~s$^{-1}$. Using this, according to
our models, $T_{\rm b}\Delta\Omega$ is between $10^{10}$ to
$10^{11}$~K, while the observed lower limit for $T_{\rm b} \approx
10^{12.4}$~K. This indicates a beaming solid angle of $\approx
10^{-3}$~sr.

\section{Discussion}

 The non-LTE method used here to model the shape of the 22~GHz H$_2$O
maser spectrum was previously used to determine the circular
polarization of the maser in the presence of a magnetic field
(V02). There the non-LTE modeling method, is compared with a standard
LTE method which is described in Vlemmings et al. (2001). The non-LTE
approach was developed to analyze circular polarization spectra and
successfully reproduced the distribution over magnetic components. In
principle, similar physical processes are important for understanding
the line shapes of the H$_2$O maser blend and we apply the same
analysis here. Recently, Watson et al. (2002) have used a similar
method to examine the line shapes of interstellar H$_2$O masers,
although their analysis has been based on only a two-level maser
transition, thus only 1 hyperfine component is included. Instead of
using the skewness of the maser profile, they examine the second order
deviations of Gaussian symmetry, the 'kurtosis'. They find that the
interstellar H$_2$O masers are found in gas with temperatures greater
than $1200$~K, for which the influence of multiple hyperfine
components is thought to be small. Since the gas in CSEs is at a lower
temperature (e.g. Goldreich \& Scoville 1976), the effect of the other
hyperfine components is larger, as is indeed seen our results.

 The large spread in the observed skewness parameter indicates that
velocity gradients along the maser path play a large role in
determining the shape of the H$_2$O maser line profiles. $35\%$ of the
maser features have a positive skewness, which only occurs for masers
that are starting to saturate and which have a
relatively large negative $\Delta V_{\rm m}$.

The observed $\Delta v \propto (T_{\rm b})^{-\alpha}$ dependence, with different
values of $\alpha$ ($ \alpha \approx -0.3$ for S~Per and $0.3$ for
VY~CMa), can be explained by the relation between beaming angle
$\Delta\Omega$ and maser optical depth $|\tau|$. As noted before, in the
completely unsaturated case $\Delta\Omega \propto |\tau|^{-1}$, while
this increases to $\Delta\Omega \propto |\tau|^{-2}$ in the saturated
case. Additionally, in the unsaturated regime the amplification is
exponential, so $T_{\rm b} \propto e^{|\tau|}$, while in the saturated
regime the amplification becomes linear and $T_{\rm b} \propto |\tau|$.
This means that, while unsaturated, the beaming angle $\Delta\Omega
\propto ({\rm ln}~T_{\rm b})^{-1}$, and when saturated $\Delta\Omega
\propto T_{\rm b}^{-2}$.  Meanwhile, our models in Fig.~\ref{width}
show, that before the maser starts re-broadening the line width is
almost constant with increasing $T_{\rm b}\Delta\Omega$. after
re-broadening, when saturation occurs, the line widths becomes
approximately proportional to $(T_{\rm b}\Delta\Omega)^{0.5}$. Using
the results for the variation of $\Delta\Omega$ with $T_b$ we then
find that for unsaturated masers $\Delta v \propto (T_{\rm b})^{0.7}$, while the
line width of saturated masers will be proportional to $(T_{\rm b})^{-0.5}$.
The value of $\alpha$ can thus increase from $\approx -0.7$ in the
mostly unsaturated regime to $0.5$ in the saturated regime. This seems
to indicate that the masers around S~Per are still mostly unsaturated,
while those around VY~CMa are close to becoming saturated. However,
the presence of velocity shifts along the maser path complicates
this straightforward explanation. For example, if the high brightness
maser features trace the lower velocity shifts because a strong
gradient disrupts the maser growth, we can also observe a decrease of
the line width with increasing brightness. Additionally, velocity
shifts will introduce a big spread in the relation between line
width and maser brightness.

 An additional uncertainty in the analysis is introduced by
assuming a linear maser model. While astrophysical masers, and
especially H$_2$O masers, are thought to be highly elongated, the
actual three-dimensional shape of the maser will affect the maser line
width and shape as described by Elitzur (1994). There it was shown
that the linear maser model is sufficient to describe the maser
profile out to $\sim~2 v_{\rm th}$ for spherical and $\sim~3$--$5 v_{\rm
th}$ for elongated cylindrical masers. However, crossing maser beams
will suppress the emission in the outer wings of the maser
profile. This can cause narrowing of the maser line to be observed
even for saturated masers. In the case of the objects studied in this
paper, the effect is expected to be small or even negligible, as the
suppression of the maser profile wings only becomes significant when
the maser becomes saturated. Then, as shown in Elitzur (1994), the
suppression effects on the profile are only noticeable when the dynamic
range of the spectrum is large. For a filamentary maser, the dynamic
range has to be higher than $\approx 7\times10^3$, while the largest
dynamic range for our maser sources is $\approx 7\times10^2$ for the
strongest maser feature of VY~CMa. If the masers were purely spherical
however, the measured linewidth decrease for increasing maser
brightness for the VY~CMa H$_2$O masers could be partly due to this
geometric effect. As a result the estimated intrinsic thermal width
and the beaming solid angle might be slightly underestimated.

The presence of large velocity gradients along the maser path, 
and the uncertainty in the exact maser geometry, increases the
difficulty of determining underlying thermal widths, level of
saturation and beaming angle. However, from comparison with the models
we have been able to give estimates for most of the sources presented
here. In general the thermal line width of the maser medium in our
sources is between $0.8$ and $1.0$ km~s$^{-1}$. This line width can be
caused by the actual temperature of the maser gas, as well as by
turbulent motions. The H$_2$O maser region should not have
temperatures higher than $T \approx 1000$ K (Neufeld \& Melnick 1990),
which implies that the line widths due to the thermal motions should
not be higher than $\approx 1.5$ km~s$^{-1}$. This is consistent with
our observations, which indicate a temperature of $T \approx 400 -
600$~K. The line width due to turbulent motions should then be of the
same order of magnitude or less. This is significantly less than the
turbulent width determined by Colomer et al.(2000), who give a width
of the order of $2-4$ km~s$^{-1}$. But as discussed above they found
larger line widths, possibly due to the fact that the VLA was not able
to resolve individual maser features.

Similar to earlier observations (e.g. Spencer et al. 1979), our
observations suggest that the circumstellar H$_2$O masers are either
unsaturated or only slightly saturated, with intrinsic brightness
temperatures between $T_{\rm b} = 10^{10} - 10^{11}$ K. From the
models we can estimate the typical beaming angle to be between
$\Delta\Omega = 10^{-2} - 10^{-3}$~sr, similar to values found before
(e.g. Nedoluha \& Watson 1991).

\section{Conclusions}

 A detailed analysis of the line profiles of 22~GHz H$_2$O masers can
 be a valuable tool to estimate saturation, thermal velocity width and
 velocity shift along the maser path in the maser region. Because the
 22~GHz transition consists of multiple hyperfine components, the
 maser profile shows clear deviations from symmetry at various stages
 of saturation. Unfortunately, the effect of velocity shifts along the
 maser path, can not easily be detached from the effects of hyperfine
 contributions. Additionally, for VY~CMa, the actual unknown maser
 geometry introduces another level of complexity.

However, we have been able to obtain some estimates on the thermal
line widths and beaming angles on the H$_2$O masers around a sample of
4 late-type stars. It is clear that velocity shifts along the maser
path of up to $1.25$ km~s$^{-1}$ play an important role, and will have to be
taken into account for further comparisons between models and
observations. The thermal line widths in the maser regions are found
to be mostly around $1.0$ km~s$^{-1}$, indicating temperatures of
approximately $400$~K in the masing region. Most masers are thought to
be only slightly saturated at the most, with $T_{\rm b}\Delta\Omega
\la 10^{11}$ K. Beaming solid angles for the circumstellar H$_2$O
masers are found to be typically between $10^{-2}$ and $10^{-3}$~sr.

{\it acknowledgments:} WV thanks the Niels Stensen Foundation for
 partly supporting his stay at Cornell University.

\end{document}